\documentclass{article}
\usepackage{spconf,amsmath,graphicx,amssymb,amsfonts,bm,algorithm,algorithmic,booktabs}
\usepackage[dvipsnames]{xcolor}
\usepackage[bibstyle=ieee,citestyle=numeric]{biblatex}
\DeclareMathOperator*{\argmax}{argmax}
\DeclareMathOperator*{\topK}{topK}
\DeclareMathOperator*{\softmax}{softmax}

\title{Jointly Learning Selection Matrices for Transmitters, Receivers and Fourier Coefficients in Multichannel Imaging}
\name{Han Wang$^{1,2}$, Yiming Zhou$^{1,3}$, Eduardo Pérez$^{1,2}$, Florian Römer$^1$\thanks{This work was partially supported by the Fraunhofer Internal Programs under Grant No. Attract 025-601128, as well as the  Thuringian Ministry of Economic Affairs, Science and Digital Society (TMWWDG).}}
\address{$^1$Fraunhofer Institute for Nondestructive Testing IZFP, Saarbrücken, Germany \\
$^2$Technische Universität Ilmenau, Ilmenau, Germany \\
$^3$Hochschule für Technik und Wirtschaft des Saarlandes, Saarbrücken, Germany}
\begin{document}
%
\maketitle
\begin{abstract}
Strategic subsampling has become a focal point due to its effectiveness in compressing data, particularly in the Full Matrix Capture (FMC) approach in ultrasonic imaging. This paper introduces the Joint Deep Probabilistic Subsampling (J-DPS) method, which aims to learn optimal selection matrices simultaneously for transmitters, receivers, and Fourier coefficients. This task-based algorithm is realized by introducing a specialized measurement model and integrating a customized Complex Learned FISTA (CL-FISTA) network. We propose a parallel network architecture, partitioned into three segments corresponding to the three matrices, all working toward a shared optimization objective with adjustable loss allocation. A synthetic dataset is designed to reflect practical scenarios, and we provide quantitative comparisons with a traditional CRB-based algorithm, standard DPS, and J-DPS.
\end{abstract}
\section{Introduction}
\label{sec:intro}
The design of subsampling patterns has gained substantial attention in many localization, detection, and imaging tasks, encompassing aspects like optimal sensor placement \cite{ospreview} and sparse array design \cite{ramalli2022design}. Notably, reducing transmissions or receptions has become crucial in various fields, e.g. in Radar \cite{roberts2011sparse, rocca2016unconventional, mulleti2020fast, diamantaras2021sparse}, in medical imaging \cite{ramalli2022design, cohen2020sparse, dps}, and in ultrasonic testing \cite{moreau2009ultrasonic, piedade2022minimum}. In the case of phased array ultrasonic imaging, Full Matrix Capture (FMC) stands out by capturing all transmitter-receiver pairs \cite{fmc}. Given its data-intensive nature, FMC demands extensive time and introduces software and hardware complexities. Consequently, exploring optimal subsampling patterns, particularly for FMC, is of paramount importance.

The data volume associated with a standard FMC configuration hinges on the count of transmitters and receivers. Additionally, due to the pulse shape produced by ultrasonic transducers, the signal energy is concentrated in a known region of the spectrum, allowing compression in the frequency domain through the judicious selection of Fourier coefficients. Each of the transmission, reception, and frequency domains can be compressed based on requirements. For example, for finite-rate-of-innovation signals, deep algorithm unfolding techniques \cite{AlgUnfolding} have showcased efficacy in optimizing Fourier coefficient sampling while maintaining recovery performance \cite{mulleti2022learning}. 

Among the available compression matrix design paradigms, the softmax neural network demonstrates promise, as evidenced by the Learn-2-Select (L2S) \cite{diamantaras2021sparse} and Deep Probabilistic Subsampling (DPS) \cite{dps} frameworks. However, they are constrained to data compression within a single domain. As they are structure agnostic, concurrent optimization along multiple domains increases the number of trainable parameters proportionally to the product of the sizes of the individual domains, making training challenging.

In this paper, we present an efficient neural network architecture comprising parallel layers that perform structured subsampling on FMC measurement data simultaneously across three axes with a largely reduced number of trainable parameters. In the context of model-based deep learning \cite{model2dl, mbdl}, we utilize a specialized Complex-Valued Neural Network (CVNN) \cite{DeepCN} for implementing a Complex Learned FISTA (CL-FISTA), thus ensuring superior performance in sparse signal recovery. Subsequently, we quantitatively contrast the Joint DPS (J-DPS) approach with the standard DPS and our prior CRB-based algorithm in terms of reconstructed image quality.

\section{Measurement Model}
\label{sec:1}
\begin{figure}[t]
  \centering
  \includegraphics[width=8cm]{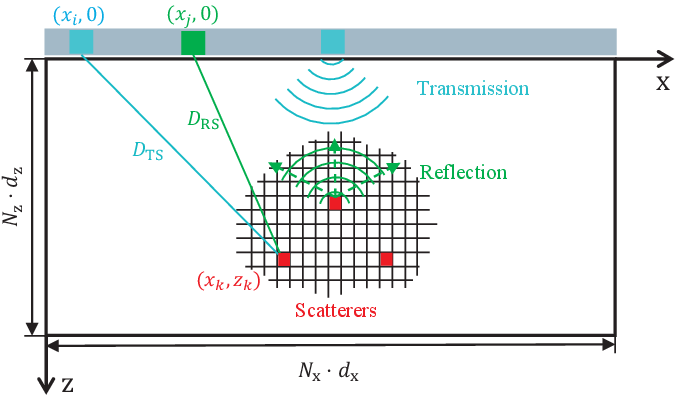}
  \caption{Illustration of a 2-dimensional measurement design featuring a specimen incorporating $K=3$ scatterers. The observed region spans dimensions of $N_{\rm{z}}\cdot d_{\rm{z}}$ in depth and $N_{\rm{x}}\cdot d_{\rm{x}}$ in length. The propagation path is the sum of the distances $D_{\rm{TS}}$ between the transmitter and scatterer and $D_{\rm{RS}}$ between the receiver and scatterer. The positional coordinates of the $i$-th transmitter, $j$-th receiver, and $k$-th scatterer are defined as $(x_i, 0)$, $(x_j, 0)$, and $(x_k, z_k)$, respectively.}
  \label{fig.FMC_measure}
\end{figure}

Although the present work focuses on ultrasound imaging, the model and methods are applicable to most multichannel nearfield imaging applications with high relative bandwidth. In our setup, we consider FMC with a uniform linear array (ULA) comprising $N_{\rm{C}}$ independent pulsing and receiving transducers, denoting the number of transmitters, receivers, and Fourier coefficients as $N_{\rm{T}}$, $N_{\rm{R}}$, and $N_{\rm{F}}$, respectively. We outline a simulation scenario illustrated in Fig. \ref{fig.FMC_measure}, where the probe is positioned on the specimen's surface, and the covered region aligns with the array's width. To effectively characterize scatterers, we discretize the specimen into ($N_{\rm{z}}\times N_{\rm{x}}$) pixels, each sized ($d_{\rm{z}}\times d_{\rm{x}}$), indicating potential scatterer locations with scattering amplitudes corresponding to their sizes.

Simulated transmitted pulses adhere to Gaussian echo profiles, while the measurement data is represented as a weighted aggregation of pulse echoes. The time delays are contingent on the geometrical interplay among the transmitters, receivers, and scatterers within the specimen. The measurement acquired from the $i$-th transmitter and $j$-th receiver can be succinctly formulated in the frequency domain as follows:
\begin{equation} \label{eq_model}
  b_{i, j}(f) = \sum_{k=1}^{K}\frac{a_k}{2} \sqrt{\frac{\pi}{\alpha}}e^{-\frac{\pi^2}{\alpha}(f-f_{\rm c})^2+j(\phi-\phi_k)-2j\pi f\tau_{k, i, j}}
\end{equation}
Here, $K$ signifies the count of isolated scatterers within the specimen, with $a_k e^{-j\phi_k}$ denoting the complex amplitude of each scatterer. The transducer's emitted pulse shape is characterized by the following parameters: a bandwidth factor $\alpha=(4.47 \times 10^6 \text{ Hz})^2$, center frequency $f_{\rm c}=4.5\ {\rm MHz}$, and phase $\phi=3 \pi /4$. The time delay $\tau$, also known as Time of Flight (ToF), is decided by the propagation path ($D_{\rm{TS}}+D_{\rm{RS}}$) and the ultrasound velocity ($c_0=6400\ {\rm{m/s}}$):
\begin{equation} \label{eq_delay}
  \tau_{k, i, j} = \frac{\sqrt{(x_k-x_i)^2+z_k^2}+\sqrt{(x_k-x_j)^2+z_k^2}}{c_{\rm 0}} 
\end{equation} 

With a sampling rate of $f_{\rm{s}}=40\ \rm{MHz}$, $N_{\rm{F}}$ Fourier coefficients $\bm{b}_{i,j} \in \mathbb{C}^{N_{\rm{F}}}$ are obtained for each single channel measurement $b_{i,j}(f)$. The FMC procedure produces a collection of $N_{\rm{R}} \cdot N_{\rm{T}}$ vectors which are stacked into the noiseless measurement data vector $\mathbf{b}\in \mathbb{C}^{N_{\rm{F}}N_{\rm{R}}N_{\rm{T}}}$. The specimen is represented by $\mathbf{x}\in \mathbb{C}^{N_{\rm{z}}\times N_{\rm{x}}}$ with scattering amplitude entries, vectorized into $\mathbf{x}\in \mathbb{C}^{N_{\rm{z}}N_{\rm{x}}}$. In the presence of noise, the linear FMC measurement model becomes:
\begin{equation}
  \mathbf{y}=\mathbf{b+n}=\mathbf{Ax+n}
\end{equation}
where $\mathbf{n}$ is circularly symmetric AWGN following $\mathcal{CN}(\boldsymbol{0}, \boldsymbol{\Sigma})$, while the measurement matrix $\mathbf{A}\in \mathbb{C}^{N_{\rm{F}}N_{\rm{R}}N_{\rm{T}}\times N_{\rm{z}}N_{\rm{x}}}$ transforms the specimen vector into measurement data based on sampled versions of \eqref{eq_model} and \eqref{eq_delay}, i.e. each column in $\mathbf{A}$ represents the expected FMC dataset that would be measured given a defect at the corresponding coordinate, represented by $\mathbf{x}$.

\section{Selection Matrices Design}
\label{sec:2}

In this work, we apply compressed sensing to the FMC measurement process via structured subsampling. We compress data by specifying active transmitters $M_{\rm{T}}$, receivers $M_{\rm{R}}$, and Fourier coefficients $M_{\rm{F}}$, represented by selection matrices: $\mathbf{S}_{\rm{T}}\in \mathbb{R}^{M_{\rm{T}}\times N_{\rm{T}}}$, $\mathbf{S}_{\rm{R}}\in \mathbb{R}^{M_{\rm{R}}\times N_{\rm{R}}}$, and $\mathbf{S}_{\rm{F}}\in \mathbb{R}^{M_{\rm{F}}\times N_{\rm{F}}}$. These matrices follow a row-wise one-hot structure, and the subsampled measurement data $\mathbf{y}_{\rm{s}}\in \mathbb{C}^{M_{\rm{F}}M_{\rm{R}}M_{\rm{T}}}$ is expressed as:
\begin{equation}
  \mathbf{y}_{\rm{s}}=(\mathbf{S}_{\rm{T}}\otimes \mathbf{S}_{\rm{R}}\otimes \mathbf{S}_{\rm{F}}) \mathbf{y}=\mathbf{S} \mathbf{A x} +\mathbf{n} \label{subsampling_measurement}
\end{equation}
where $\otimes$ denotes the Kronecker product and the three matrices can be unified into a single matrix $\mathbf{S}\in \mathbb{R}^{M_{\rm{F}}M_{\rm{R}}M_{\rm{T}}\times N_{\rm{F}}N_{\rm{R}}N_{\rm{T}}}$.

\subsection{CRB-based Exhaustive Search Algorithm}
The selection matrix can be constructed in an estimator-agnostic fashion by considering the Cramér-Rao Bound (CRB) of the parameters of interest. As a baseline to compare the present contribution against, we design the structured selection matrix by first choosing $\mathbf{S}_{\rm{F}}$ such that $M_{\rm{F}}$ Fourier coefficients are chosen around the center frequency $f_c$ of the pulse shape. Then, a combinatorial minmax problem on the trace of the CRB is solved through exhaustive search. We refer the readers to Algorithm 1, Section 6 of \cite{mdpi} for a detailed explanation of the approach.

\subsection{Deep Probabilistic Subsampling}
We construct the selection matrix by treating each row as a sample from a categorical distribution. To address the challenge posed by discrete variables within neural networks, a conventional approach is the Gumbel-Softmax reparameterization trick\cite{jang2017categorical, maddison2017concrete}. An unnormalized log-probability (logits) vector $\bm{\theta}$ is mapped to a one-hot row vector $\mathbf{s}$, where each $\theta_i$ signifies the probability of $s_i$ being 1. 

Instead of using a separate vector $\bm{\theta}_i$ per row of the subsampling matrix as in \cite{diamantaras2021sparse}, we note that the subsampling procedure can be carried out through elementwise multiplication with a single binary vector \cite{IUS, EUSIPCO}. Then the derivation of selection vector, for example $\tilde{\mathbf{s}}_{\rm{T}}$, and the subsampled measurement model (\ref{subsampling_measurement}) can be rewritten as:
\begin{equation}
    \tilde{\mathbf{s}}_{\rm{T}}=\topK\left( \frac{\exp((\bm{\theta}_{\rm{T}}+\bm{g})/\gamma)}{\sum \exp((\bm{\theta}_{\rm{T}}+\bm{g})/\gamma)} \right)
    \label{eq. topk}
\end{equation}
\begin{equation}
    \mathbf{y}_{\rm{s}}=(\tilde{\mathbf{s}}_{\rm{T}}\otimes \tilde{\mathbf{s}}_{\rm{R}}\otimes \tilde{\mathbf{s}}_{\rm{F}})\odot \mathbf{y}=\tilde{\mathbf{s}}\odot \mathbf{A x} +\mathbf{n} \label{sub_meas_new}
\end{equation}
Here, $\bm{g}$ represents a noise vector drawn from $\rm{Gumbel}(0, 1)$, and $\gamma$ serves as the softmax temperature. Then $\tilde{\mathbf{s}}\in \mathbb{R}^{ N_{\rm{F}}N_{\rm{R}}N_{\rm{T}}}$ has $(M_{\rm{F}}M_{\rm{R}}M_{\rm{T}})$ active elements and the Gumbel top K trick results in a single vector $\bm{\theta}$ of log probabilities. 

For the FMC case presented in (\ref{subsampling_measurement}), our interest lies in jointly optimizing three selection vectors, rather than directly training for $\tilde{\mathbf{s}}$, owing to two key reasons. Firstly, this further reduces the trainable parameter count to $(N_{\rm{F}} + N_{\rm{R}} + N_{\rm{T}})$ due to the repeating structure of the subsampling pattern. Secondly, the resulting structured subsampling pattern can be more practically implemented in hardware.

\section{Joint DPS Approach}
\label{sec:4}

\subsection{Complex-Valued Learned FISTA}
As a task-based algorithm, our aim is to achieve optimal selection matrices while maintaining exceptional sparse signal recovery performance. A commonly employed strategy to incorporate signal recovery into subsampling pattern optimization is to concatenate a Learned ISTA (LISTA) network \cite{dps, mulleti2022learning, LISTA}. In order to accommodate complex-valued data in the frequency domain, we modify the standard LISTA into a Complex-Valued Neural Network (CVNN) \cite{DeepCN, CVNN}. Among the available approaches, we utilize two distinct real-valued linear layers, each responsible for the real and imaginary components of the input. This results in fewer trainable parameters compared to stacking the real and imaginary parts and training a larger weight matrix.

We introduce a trainable softshrink parameter for each layer, while incorporating Nesterov acceleration into each iteration, resulting in a Complex Learned FISTA (CL-FISTA) approach. We note that, after unfolding, this form of momentum becomes part of the network architecture and is unrelated to any momentum employed by the optimizer with which the network parameters are trained. After meticulous hyperparameter tuning, we have chosen the L1 norm as the cost function due to its ability to yield high-quality images with precise point scatterer localization.

\subsection{Joint DPS Network Architecture}
\begin{algorithm}[ht]
  \renewcommand{\algorithmicrequire}{\textbf{Input:}}
  \renewcommand{\algorithmicensure}{\textbf{Output:}}
  \caption{Joint DPS with CL-FISTA}
  \label{alg1}
  \begin{algorithmic}[1]
    \REQUIRE Training data pairs: $(\mathbf{x, \, y})$, measurement model: $\mathbf{A}$, largest eigenvalue of $\mathbf{A}^{\rm{H}}\mathbf{A}$: $\mu$, parameters of full- and subsampling: $N_{\rm{T}}$, $N_{\rm{R}}$, $N_{\rm{F}}$, $M_{\rm{T}}$, $M_{\rm{R}}$, $M_{\rm{F}}$, number of CL-FISTA layers: $N_{\text{layer}}$, softshrink parameters vector: $\bm{\delta}\in \mathbb{R}^{N_{\text{layer}}}$
    \STATE \textbf{Initialization:} \\
      - Weight matrices of CL-FISTA: $\mathbf{W}=\mathbf{I}-\mu\mathbf{A}^{\rm{H}}\mathbf{A}$ and $\mathbf{V}=\mu\mathbf{A}^{\rm{H}}$ \\
      - Logits vectors: $\bm{\theta}_{\rm{T}},\bm{\theta}_{\rm{R}},\bm{\theta}_{\rm{F}}\sim \mathcal{U}(0, 0.1)$ \\
      - Auxiliary and real reconstructed signal: $\mathbf{x}_{\text{au}}=\hat{\mathbf{x}}=\bm{0}$ \\
      - Momentum parameter: $\eta = 1$ \\
      - Initial and end temperature: $\gamma_{\text{init}}=5$ and $\gamma_{\text{end}}=0.5$ \\
      - Temperature decay rate: $\triangle\gamma=\frac{\gamma_{\text{init}}-\gamma_{\text{end}}}{N_{\text{iter}}}$
    \FOR{$i=1$ to $N_{\rm iter}$}
      \STATE Current temperature: $\gamma=\gamma_{\text{init}}-(N_{\rm iter}-1)\cdot \triangle\gamma$ \\
      \STATE Parallelly compute selection vectors: \\
        - $\tilde{\mathbf{s}}_{\rm{T}}=\topK\left( \softmax(\bm{\theta}_{\rm{T}},\, \bm{g},\, \lambda )\right)$ \\
        - $\tilde{\mathbf{s}}_{\rm{R}}=\topK\left( \softmax(\bm{\theta}_{\rm{R}},\, \bm{g},\, \lambda )\right)$ \\
        - $\tilde{\mathbf{s}}_{\rm{F}}=\topK\left( \softmax(\bm{\theta}_{\rm{F}},\, \bm{g},\, \lambda )\right)$ \\
      \STATE Unified selection vector: $\tilde{\mathbf{s}}=\tilde{\mathbf{s}}_{\rm{T}}\otimes \tilde{\mathbf{s}}_{\rm{R}}\otimes \tilde{\mathbf{s}}_{\rm{F}}$ \\
      \STATE Subsample the measurements: $\mathbf{y}_{\rm{s}}=\tilde{\mathbf{s}}\odot \mathbf{y}$ \\
      \FOR{$j=1$ to $N_{\text{layer}}$}
        \STATE $\hat{\mathbf{x}}_{\text{old}}=\hat{\mathbf{x}}$ \\
        \STATE $\hat{\mathbf{x}}=h_{\delta_j}(\mathbf{Wx}_{\text{au}}+\mathbf{Vy}_{\rm{s}})$ \\
        \STATE $\eta_{\text{old}}=\eta$ \\
        \STATE $\eta=\frac{1+\sqrt{1+4\eta_{\text{old}}^2}}{2}$ \\
        \STATE $\mathbf{x}_{\text{au}}=\hat{\mathbf{x}}+\frac{\eta_{\text{old}}-1}{\eta}(\hat{\mathbf{x}}-\hat{\mathbf{x}}_{\text{old}})$ \\
      \ENDFOR
      \STATE Compute loss: $\mathcal{L}=\Vert \mathbf{x}-\hat{\mathbf{x}} \Vert_1=\Vert \mathbf{x}-\text{CL-FISTA}(\mathbf{y}_{\mathrm s}) \Vert_1$\\
      \STATE Use Adam optimizer to update logits and CL-FISTA parameters by minimizing $\mathcal{L}$\\
    \ENDFOR
    \ENSURE  Trained logits vectors $\bm{\theta}_{\rm{T}}$, $\bm{\theta}_{\rm{R}}$ and $\bm{\theta}_{\rm{F}}$, CL-FISTA network parameters $\mathbf{W}$, $\mathbf{V}$ and $\bm{\delta}$.
  \end{algorithmic}  
\end{algorithm}

\begin{figure}[h]
    \centering
    \includegraphics[width=8cm]{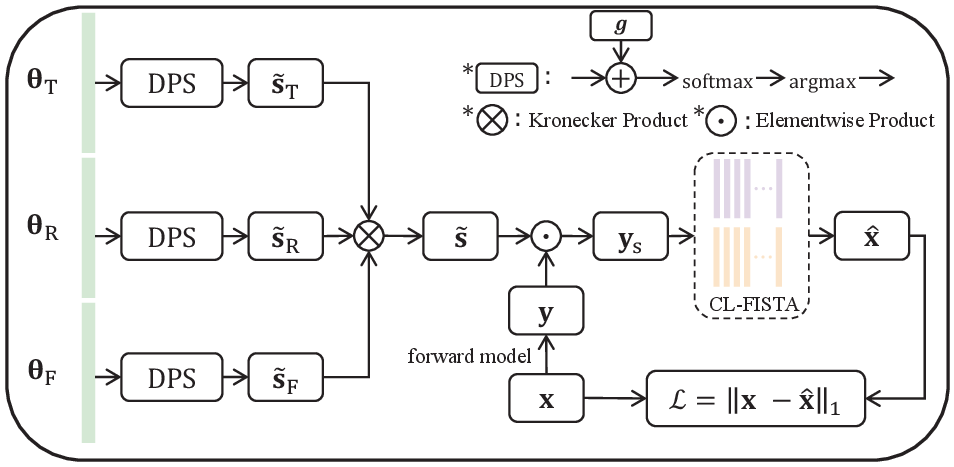}
    \caption{Network architecture of J-DPS.}
    \label{fig.J-DPS_network}
\end{figure}

The operational procedure of Joint DPS (J-DPS) is illustrated in Algorithm \ref{alg1}. The network architecture, as shown in Fig. \ref{fig.J-DPS_network}, is composed of two primary components: a parallel network responsible for updating three selection vectors and the CL-FISTA network for sparse signal recovery.

Each parallel network is designed with an identical architecture, and they operate independently on the three selection vectors, allowing concurrent training. The logits vector is represented by the biases of a linear layer \cite{EUSIPCO} and properly initialized as shown in Algorithm \ref{alg1}. In training, the probability of optimal selections will gradually increase, therefore, initial values with a smaller difference can accelerate the process. Subsequently, each output of the layer enters a DPS block, including the addition of Gumbel noise, as well as $\softmax$, and $\argmax$ operations. Then the three selection vectors are unified into the vector $\tilde{\mathbf{s}}$ through a Kronecker product. After using $\tilde{\mathbf{s}}$ to perform subsampling, the compressed data is passed into CL-FISTA to obtain a reconstruction.

As to the synthetic data generation, we address two practical scenarios. In the first, the specimen contains a vulnerable area prone to defects, while in the second, a region of critical significance requires heightened detection sensitivity and improved image resolution. To achieve this, scatterers with varying scattering amplitudes are concentrated within a designated area. Specifically, the specimen shown in Fig. 1 measures 8 mm in length and 6.4 mm in depth. We define a region spanning from 4 mm to 6 mm in length and 3.2 mm to 4.8 mm in depth to align with this strategy. The scatterer count in each specimen follows a discrete uniform distribution $\mathcal{U}\{1,2,3,4,5\}$, with scattering amplitudes adhering to $\mathcal{CN}(10, 5)$. 

\section{Evaluation}
\label{sec:5}

Given the spatial coverage and the sampling frequency, we require 128 time-domain samples, which translates to 65 Fourier coefficients. Initially, we apply compression by selecting 23 coefficients within a bandwidth of 1 MHz to 8 MHz centered around $f_{\rm{c}}$. For algorithmic comparison purposes, we employ a small array probe consisting of 8 transducers.

In the comparative analysis, we utilize the CRB-based algorithm, standard DPS, and J-DPS to process the same test dataset. The specific subsampling patterns are summarized in Table \ref{comparison_table}. DPS-T and DPS-F denote compression of only the transmitters and only the Fourier coefficients, respectively. Subsequently, we reconstruct the signal from the subsampled data and quantitatively assess image quality using the Mean Squared Error (MSE) metric. For the CRB-based algorithm, a standard FISTA is employed to produce images.

\begin{figure}[htb]
    \centering
    \includegraphics[width=7.7cm]{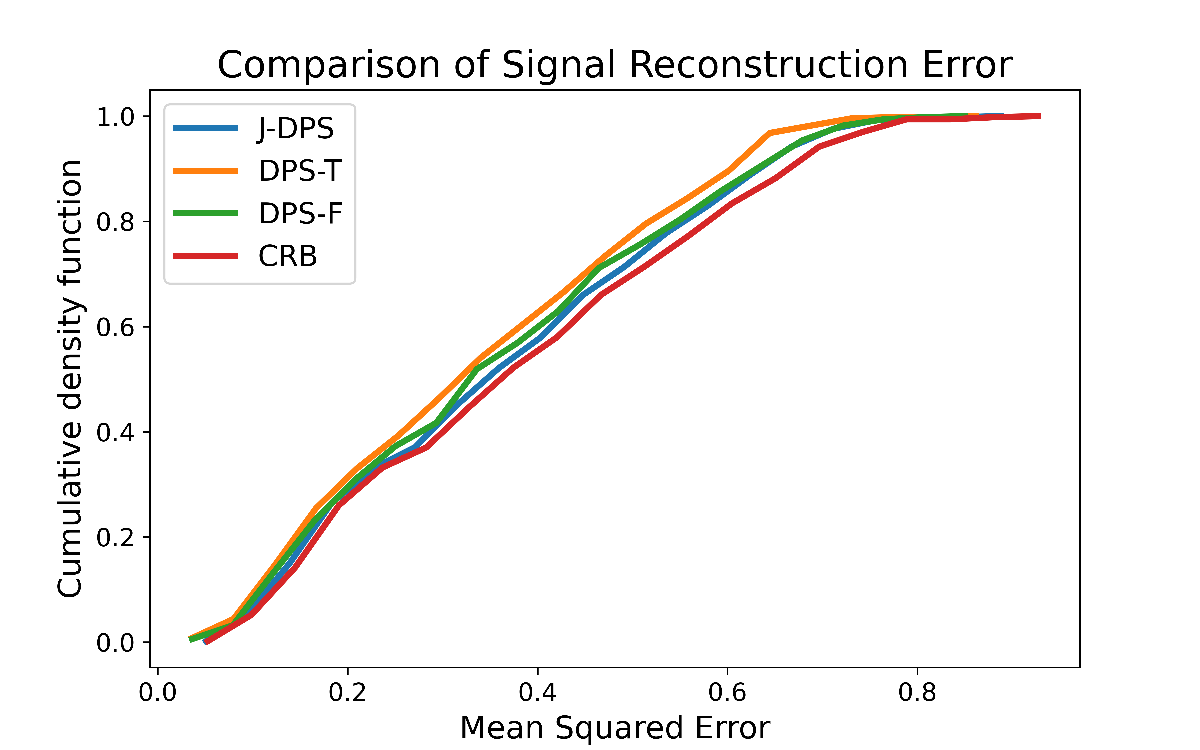}
    \caption{Comparison of reconstruction error using four subsampling techniques. Higher and shorter curves indicate smaller MSE and better image quality. J-DPS performs similarly to the other techniques, but allows greater compression ratios than DPS-T and DPS-F, and can be applied to larger problems than the CRB-based method.}
    \label{fig.cdf}
\end{figure}

To provide a statistical representation of the comparison results, we compute the Cumulative Density Function (CDF) with respect to MSE, as depicted in Figure \ref{fig.cdf}. A steep increase in the curve indicates smaller errors dominating. Notably, the J-DPS approach exhibits superior performance over the CRB-based algorithm at the same compression ratio while additionally scaling better to larger scenarios. The remaining two methods only marginally outperform J-DPS despite collecting significantly more data, while also circumventing the loss of resolution that can arise when aggressively compressing along a single axis \cite{kirchhof2021frequency}. Additionally, DPS-T outperforms DPS-F, rousing interested in methods to optimally distribute the overall compression among the available axes.

\begin{table}[htb!]
    \caption{Subsampling setup}\label{comparison_table}
    \setlength{\tabcolsep}{2mm}
    \begin{center}
    \begin{tabular}{ccccc}
    \toprule
    \textbf{Algorithm}& $M_{\rm{T}}$ & $M_{\rm{R}}$ & $M_{\rm{F}}$ & Compression Ratio \\
    \midrule
    \textbf{J-DPS}& $3$ & $4$ & $9$ & $2.5\%$ \\
    \textbf{CRB-based}& $3$ & $4$ & $9$ & $2.5\%$\\
    \textbf{DPS-T}& $3$ & $8$ & $23$ & $13.3\%$\\
    \textbf{DPS-F} & $8$ & $8$ & $9$ & $13.8\%$\\
    \bottomrule
    \end{tabular}
    \end{center}
\end{table}

\section{Conclusion}
\label{sec:print}
This paper tackles the challenge of jointly optimizing subsampling matrices for transmitters, receivers, and Fourier coefficients in phased array ultrasonic imaging. The J-DPS approach is realized by integrating a CL-FISTA network into a parallel architecture that enables simultaneous updates of the three subsampling matrices. Even with significantly reduced data sampling, our approach maintains high-quality image reconstructions. In quantitative evaluations, J-DPS outperforms the traditional CRB-based algorithm at the same compression ratio, while also being applicable to larger problems for which full search algorithms are intractable. Furthermore, standard DPS is limited in the amount of compression it can achieve due to acting on a single axis. 

Due to the high dimensionality of the forward model, training remains a time-intensive process. Additionally, performance differences when compressing along different axes were observed. Therefore, future work will consider incorporating convolutional layers into CL-FISTA to expedite the overall process \cite{sreter2018learned}, as well as adaptively resizing the selection matrices based on performance. Alternatively, the contribution of each of the axes to the loss function can be weighted, e.g. prioritizing transmitter subsampling to reduce measurement time.

\printbibliography[heading=bibnumbered, title={{REFERENCES}}]

\end{document}